\documentclass[twocolumn,aps,pra,showpacs,preprintnumbers, superscriptaddress, amsmath,amssymb]{revtex4}

\usepackage{stmaryrd} 
\usepackage{mathtools} % for symbols like := and =:
\usepackage{stmaryrd} % for variable-sized delimiters
\newcommand {\sign}{{\rm sign}\,}
\newcommand {\Tr}{{\rm Tr}\,}

\usepackage{amsmath}
\usepackage{amsfonts}

\usepackage{amssymb}
\usepackage{graphicx}%

\begin{document}

\author{Andre G. Campos}
\email{agontijo@princeton.edu}
%\affiliation{Department of Chemistry, Princeton University, Princeton, NJ 08544, USA} 

\author{Renan Cabrera}
%\email{dbondar@princeton.edu}
%\affiliation{Department of Chemistry, Princeton University, Princeton, NJ 08544, USA} 

\author{Denys I. Bondar}
%\email{rcabrera@princeton.edu}
%\affiliation{Department of Chemistry, Princeton University, Princeton, NJ 08544, USA} 

\author{Herschel A. Rabitz}
%\email{hrabitz@princeton.edu}
\affiliation{Princeton University, Princeton, NJ 08544, USA} 
\title{Excess of positrons in cosmic rays: A Lindbladian model of quantum electrodynamics}

\date{\today}

\begin{abstract}
The fraction of positrons and electrons in cosmic rays recently observed on the International Space Station unveiled an unexpected excess of the positrons, undermining the current foundations of cosmic rays sources. We provide a quantum electrodynamics phenomenological model explaining the observed data. This model incorporates electroproduction, in which cosmic ray electrons decelerating in the interstellar medium emit photons that turn into electron-positron pairs. These findings not only advance our knowledge of cosmic ray physics, but also pave the way for computationally efficient formulations of quantum electrodynamics, critically needed in physics and chemistry.
\end{abstract}

\maketitle

\begin{figure}
  \includegraphics[scale=0.45]{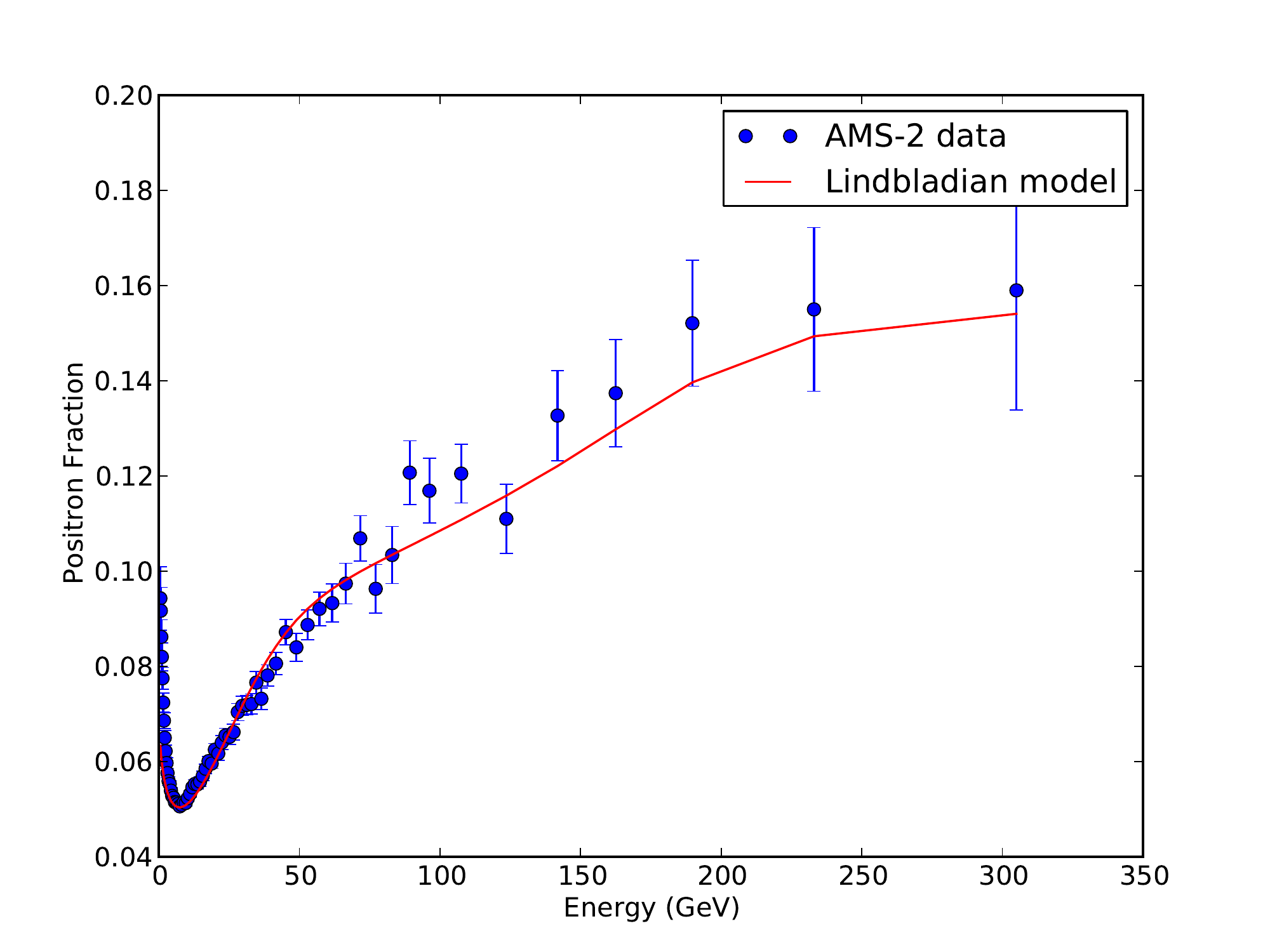}
  \caption{(color online) Comparison between the proposed model for electroproduction and the measured positron fraction in cosmic rays.} 
     \label{fig2}
\end{figure}

\emph{Introduction.} Cosmic ray electrons are primarily believed to be generated form quasars, micro quasars, and supernovas \cite{MoskalenkoTheAstrophysicalJ493-694(1998)}. Collisions of cosmic ray hadrons with the interstellar medium yield positrons, which are characterized by a monotonic decrease with energy \cite{FengPhysicsLettersB728-250(2014),CholisPhysRevD88-023013(2013),KoppPhysRevD88-076013(2013),LavallePhysRevD90-081301(2014),CholisPhysRevD89-043013(2014)}. However, this explanation is challenged by recent measurements of cosmic ray electron and positron fluxes with energies ranging from $0.5$ GeV to $1$ TeV, detected
 by the Alpha Magnetic Spectrometer (AMS) on the International Space Station \cite{Aguilar1,Aguilar2,Accardo,Aguilar3} (see Fig. \ref{fig2}). These observations revealed an inexplicable rise of the positron fraction at higher energies, which is a subject of heated debate \cite{SerpicoPhysRevD79-021302(2009),AdrianiNature458-607(2009),SerpicoAstroParticlePhysics39-40-2(2012),BlumPhysRevLett111-211101(2013),CholisPhysRevD88-023013(2013),KoppPhysRevD88-076013(2013),LavallePhysRevD90-081301(2014),CholisPhysRevD89-043013(2014)}. In particular, an annihilation of dark matter has been entertained as the unaccounted source of positrons \cite{AdrianiNature458-607(2009),SerpicoPhysRevD79-021302(2009),CholisPhysRevD88-023013(2013),KoppPhysRevD88-076013(2013),LavallePhysRevD90-081301(2014),CholisPhysRevD89-043013(2014),IbarraPhysRevD89-063539(2014)}. Less speculative sources have also been put forth \cite{BlumPhysRevLett111-211101(2013),MertschPhysRevD90-061301(R)(2014)}, but these explanations were found to be inconsistent with the observed boron-to-carbon ratio \cite{AdrianiNature458-607(2009),CholisPhysRevD89-043013(2014)}.

In this Article, we show that the positron excess can arise as a result of the cosmic ray electrons directly interacting with the interstellar medium. In fact, ultra-relativistic electrons injected into a medium produce electron-positron pairs through the process of electroproduction \cite{BaierJETPLetters88-80(2008),BaierPhysicsLettersA373-1874(2009),EsbergPhysRevD82-072002(2010)}. One possible channel of such a reaction is as follows: Under the influence of the external Coulomb field of a nucleus, the incident electron emits an energetic gamma quant that subsequently decays into a positron-electron pair. We propose a relativistic Lindbladian model for electroproduction that not only quantitatively describes the observed positron excess (see Fig. \ref{fig2}), but also predicts a plateau for the positron fraction in the high energy limit. 

\emph{The Lindbladian model.} In order to derive the master equation, we employ the paradigm of \textit{Operational Dynamic Modeling} (ODM) \cite{Denys_ODM, Denys}, designed to deduce equations of motion from the dynamics of observable averages, which are supplied in the form of Ehrenfest-like relations. Electroproduction is modeled as a dissipative interaction of an energetic incoming electron with a medium. Since velocities of the incident particles are ultra-relativistic, a dissipative model for the Dirac fermion is required. The starting point of ODM are the Ehrenfest theorems for the coordinate ($\hat{x}^\mu$) and momentum ($\hat{p}^\mu$), which in the case of the free-particle Dirac equation (i.e., non-dissipative relativistic dynamics) read \cite{Hestenes1973,Hestenes1975,Renan_first}
\begin{align}\label{FreeDiracEhrenfest}
	\frac{d}{ds}\langle\hat{x}^\mu\rangle=c\langle\gamma^0\gamma^\mu\rangle,\qquad
	\frac{d}{ds}\langle\hat{p}^\mu\rangle=0. 
\end{align}
Here, $c$ is the speed of light, $\gamma^\mu$ with $\mu=0,1,2,3$ are the Dirac matrices, and $s$ is the proper time. In non-relativity the momentum and velocity operators are given by $\hat{p}^k$ and $\hat{p}^k/m$, $k=1,2,3$, respectively. However, because a relativistic particle cannot move faster than light, the momentum and velocity operators have fundamentally different forms: $\hat{p}^k$ and $c\gamma^0\gamma^k$, respectively. The former is unbounded, while the latter is bound by the speed of light \cite{Greiner_book2}. Note that Eq. (\ref{FreeDiracEhrenfest}) is covariant. 

The magnitude of a dissipative force of friction is proportional to the particle's velocity relative to the medium, slowing the particle down. Therefore, the inclusion of the friction leads to the following Ehrenfest theorems
\begin{align}
	\frac{d}{dt}\langle\hat{x}^k\rangle=c\langle\gamma^0\gamma^k\rangle, \qquad
	\frac{d}{dt}\langle\hat{p}^k\rangle= -2\sigma mc\langle\gamma^0\gamma^k\rangle,
	\label{eq3}
\end{align}
where $\sigma \geq0$ is a friction coefficient phenomenologically characterizing the medium. Since the velocity is measured with respect to the medium, Eqs. (\ref{eq3}) are non-covariant, and $t$ is the time in the frame of reference attached to the medium.

The Ehrenfest theorems (\ref{eq3}) can be reached from many other perspectives. For example,
in an analogy with Ohm's law, microscopically describing collisions \cite{Ohm}, an Ohmic environment is coupled to the particle's momentum through the current, whose average for Dirac fermions is $\langle \gamma^0\gamma^k \rangle$ \cite{Schwabl_book,Greiner_book}. Additionally, a particle is coupled to the external photon field through the velocity operator, as prescribed by the minimal coupling in quantum electrodynamics \cite{BetheSalpeter_book,Greiner_book2}. Moreover, a classical counterpart of Eq. (\ref{eq3}), where all the operators are replaced by corresponding classical quantities, is closely related to the Langevin equation for the relativistic Ornstein-Uhlenbeck process \cite{DebbaschJStatisticalPhysics88-945(1997),DebbaschJStatisticalPhysics90-1179(1998)}, describing the velocity of a Brownian particle under the influence of friction.

Equations (\ref{eq3}) are consistent with a picture of open system dynamics. Hence, the average of an observable $\hat{O}$ is defined by
\begin{align}
\langle\hat{O}\rangle=\Tr[\hat{O}\hat{\rho}],
\label{eq1}
\end{align}
where $\hat{\rho}$ is a density matrix representing the state of the system, whose evolution equations is of the Lindbladian form \cite{KossakowskiRepMathPhys3-247(1972),LindbladCommMathPhys48-119(1976),MunroPhysRevA53-2633(1996)} 
\begin{align}
	\frac{d}{dt}\hat{\rho}&=-\frac{i}{\hbar}[\hat{H},\hat{\rho}]+D[\hat{\rho}],\nonumber\\
	\hat{H}&=c\alpha^k \hat{p}_k + mc^2\gamma^0, \nonumber\\
	D[\hat{\rho}]&=\frac{\sigma}{\hbar}\left(\hat{B}\hat{\rho}\hat{B}^\dagger-\frac{1}{2}	(\hat{\rho}\hat{B}^\dagger\hat{B}+\hat{B}^\dagger\hat{B}\hat{\rho})\right),
	\label{eq4}
\end{align}
where $\alpha^k=\gamma^0\gamma^k$ and $\hat{B}$ is an unknown operator to be found following Ref. \cite{Denys}.

 Hereinafter, we will consider one spatial dimension without loss of generality. Using the relations
\begin{align}
[\hat{B}(\hat{x},\hat{p}),\hat{p}]=i\hbar\partial\hat{B}/\partial\hat{x},\quad
[\hat{B}(\hat{x},\hat{p}),\hat{x}]=-i\hbar\partial\hat{B}/\partial\hat{p},
\label{eq5}
\end{align}
that follow from the canonical commutation relation $[\hat{x},\hat{p}]=i\hbar$ and Eqs. (\ref{eq1}), (\ref{eq3}) and (\ref{eq4}), we obtain
\begin{align}
\Tr\left[\left(\hat{B}^\dagger\frac{\partial\hat{B}}{\partial\hat{p}}-\frac{\partial\hat{B}^\dagger}{\partial\hat{p}}\hat{B}\right)\hat{\rho}\right]&=0,\nonumber\\
\Tr\left[\left(\hat{B}^\dagger\frac{\partial\hat{B}}{\partial\hat{x}}-\frac{\partial\hat{B}^\dagger}{\partial\hat{x}}\hat{B}\right)\hat{\rho}\right]&=-4imc\Tr[\alpha^1\hat{\rho}].
\label{eq6}
\end{align}
We require that these identities should be valid for any state $\hat{\rho}$, hence
\begin{align}
\hat{B}^\dagger\frac{\partial\hat{B}}{\partial\hat{p}}-\frac{\partial\hat{B}^\dagger}{\partial\hat{p}}\hat{B}&=0,\nonumber\\
\hat{B}^\dagger\frac{\partial\hat{B}}{\partial\hat{x}}-\frac{\partial\hat{B}^\dagger}{\partial\hat{x}}\hat{B}&=-4imc\alpha^1.
\label{eq7}
\end{align}
These are equations for the unknown operator $\hat{B}=B(\hat{x},\hat{p})$ defined as a function of the non-commutative variables within the Weyl calculus \cite{Weyl_C1,Weyl_C2,Weyl_C3}. Performing the Weyl transform, the operator equations (\ref{eq7}) turn into a system of equations for the complex-valued matrix function $B(x,p)$ defined on the phase space
\begin{align}
B^\dagger(x,p)\star\frac{\partial B(x,p)}{\partial p}-\frac{\partial B^\dagger(x,p)}{\partial p}\star B(x,p)&=0,\nonumber\\
B^\dagger(x,p)\star\frac{\partial B(x,p)}{\partial x}-\frac{\partial B^\dagger(x,p)}{\partial x}\star B(x,p)&=-4imc\alpha^1,
\label{eq8}
\end{align}
where $\star$ denotes the Moyal product \cite{Weyl_C1,Weyl_C2,Weyl_C3}
\begin{align}\label{MoyalStar}
	\star = \exp \frac{i\hbar}{2} \left( 
		\overleftarrow{\frac{\partial}{\partial x}} \overrightarrow{\frac{\partial}{\partial p}} -
		\overleftarrow{\frac{\partial}{\partial p}} \overrightarrow{\frac{\partial}{\partial x}}
	 \right).
\end{align}
The system (\ref{eq8}) has the following exact solution
\begin{align}
B(x)=-i\sqrt{mc\lambdabar}-2\sqrt{mc/\lambdabar}~\alpha^1x,
\label{eq9}
\end{align}
where $\lambdabar$ is a constant with the dimension of $\mbox{length}$, whose physical meaning will be elucidated below. 

Even though $D[\hat{\rho}]$ is non-translationally invariant, it can be replaced by the  translationally invariant modification within the interval $[-L,L]$, 
\begin{align}
	D[\rho](x) \to \frac{1}{2L}\int_{-L}^L dx_0 D[\rho](x-x_0),
	\label{eq9a}
\end{align}
without changing the underlying Ehrenfest theorems (\ref{eq3}).

Let us find the non-relativistic limit of $B(x)$. Note that $B(x)$ is a matrix valued function of $x$. Calculating the eigenvalues of $B(x)$, denoted by $B_e$, and taking into account the shell-mass condition $mc=\sqrt{E^2/c^2-p^2}$, we arrive at
\begin{align}
	\lim_{c\rightarrow\infty}(B_e e^{-5i\pi/4})=(i\lambdabar+2x~\sign{p})\sqrt{|p|/	\lambdabar},
	\label{limit_dissipator}
\end{align}
Equation (\ref{limit_dissipator}) coincides with the Lindbladian operator for the non-relativistic quantum friction recently obtained in Ref. \cite{Denys}, which leads to the interpretation of $\lambdabar$ as the thermal de Broglie wavelength, phenomenologically characterizing the medium. 

\emph{Positron fraction in cosmic rays.} 
Using a modification of the numerical method developed in Ref. \cite{Renan}, we propagate Eqs. (\ref{eq4}), (\ref{eq9}), and (\ref{eq9a}) in the phase space, where the density matrix  $\hat{\rho}$ is represented by the Wigner function $W(t,x,p)$. To construct an initial state $W(0,x,p)$ composed only of electrons, we apply the phase-space generalization of the projection operator \cite{CamposPhysRevA90-034102(2014)} to remove positrons from a Dirac gaussian spinor with the mean momentum $p_0$ and the spatial standard deviation $a$.
Time evolution generates positrons. To separate the state of created positrons $W_p(t,x,p)$ from the total Wigner function $W(t,x,p)$, we project the electrons out from $W(t,x,p)$, as prescribed in Ref. \cite{CamposPhysRevA90-034102(2014)}. The fraction of positrons at the final time $T$ is defined as $N\int  W_p(T,x,p) dx /\int  W(T,x,p) dx$, with $N$ being a normalization constant. In the ultra-relativistic regime, which is employed here, $p$ equals to the kinetic energy. For the propagator not to run into numerical stability issues due to a wide range of energies spanning the horizontal axis in Fig. \ref{fig2}, we scale the momentum $p \to \epsilon p$. The values of $\lambdabar=400$ and $T=7.68$, in natural units, are chosen for numerical convenience and not varied during the fitting procedure. 

The five parameters $a=1/\sqrt{3.4}$, $p_0=4.8$, $\sigma=0.37$, $\epsilon=22.5891$, and $N=0.154$ were used to fit the experimental points in Fig. \ref{fig2}. It is important to note that the overall shape of the curve in Fig. \ref{fig2} is maintained for a wide range of values of the parameters. The first two ($a$ and $p_0$) define the distribution of the incident electrons, the friction coefficient $\sigma$ characterizes the interstellar medium, and the energy scaling constant $\epsilon$ is introduced for a technical purpose described above. The position of the positron fraction minimum in Fig. \ref{fig2} depends on $p_0$ and $\sigma$; whereas, the slope of the distribution at higher energies is controlled by $a$.

The developed Lindbladian model of electroproduction predicts that the positron fraction reaches a plato at high energies. The plateau seen in Fig. \ref{fig2} is attributed to the fact that the friction force, causing electroproduction, is bounded because its magnitude is proportional to the velocity limited by the speed of light. Thus, pair production saturates for sufficiently high energies. Note that the experimental data in Fig. \ref{fig2} cannot be fitted if the Ehrenfest theorems (\ref{eq3}) are replaced by
\begin{align}
	\frac{d}{dt}\langle\hat{x}^k \rangle =c\langle\gamma^0\gamma^k \rangle, \qquad
	\frac{d}{dt}\langle\hat{p}^k \rangle =-2\sigma\langle \hat{p}^k \rangle,
\end{align}
where the dissipative force is proportional to the momentum, which is unbounded. 

\emph{Conclusions.}
In this Article we develop the master equation for electroproduction, a higher-order quantum electrodynamic process resulting in creation of electron-positron pairs upon ultra-relativistic electrons interacting with the medium. This model, being in an excellent agreement with observed data \cite{Aguilar1,Aguilar2,Accardo,Aguilar3} (see Fig. \ref{fig2}), accounts for the unexpected rise of the positron fraction in cosmic rays, and thus, renders explanations involving dark matter decay unnecessary. According to our view, additional positrons are generated from cosmic ray electrons colliding with the interstellar medium leading to electron-positron pair production.

Moreover, the ideas presented in the Article serve as a major step towards computationally clean formulation of quantum electrodynamics, which is bound to open new horizons in many-body quantum electrodynamics calculations, where a discrepancy between theory and experiment has been observed \cite{chantler2012,Notermans2014}. It has been known for a long time \cite{Welton1948} that quantum electrodynamical effects can be viewed as emerging from a particle interacting with a noisy environment, modeling the vacuum. Such a point of view currently enjoys a revived interest due to the need for computationally accessible alternatives for incorporating quantum electrodynamical effects into chemical structure calculations \cite{DIRAC13}. All these approaches, nevertheless, have a limited rage of applicability because they represent \emph{ad hoc} modification of the Dirac equation, describing dynamics of closed systems. A Lindbladian master equation can give a universally applicable and inconsistency-free description of a quantum system coupled to an environment. In this regard, the presented derivation of the master equation for electroproduction lays out a roadmap to accomplish computationally efficient Lindbladian description of quantum electrodynamics.

\bibliography{bib-relativity}
\end{document}